\documentclass[12pt,twoside]{article}

\usepackage{jeep}       
\mysection{section}{\large \bf}{\thesection.~}   
\mysection{subsection}{\normalsize\bf}{\thesubsection.~}

\usepackage{a4wide}     

\usepackage{fancyheadings}

\usepackage{cite}   
\usepackage{epsf}
\usepackage{rotate}
\usepackage{subeqn}

\advance \headheight by 3.0truept       

\usepackage{times}

\newcommand{\lt}{\left}
\newcommand{\rt}{\right}
\newcommand{\ov}{\overline}

\newcommand{\eq}[1]{(\ref{#1})}

\def\openone{\leavevmode\hbox{\small1\kern-3.8pt\normalsize1}}%

\newcommand{\nn}{\nonumber \\}
\newcommand{\no}{\nonumber }
\newcommand{\fig}[1]{Fig.~\ref{#1}}
\newcommand{\tab}[1]{Tab.~\ref{#1}}

\newcommand{\imag}{{\rm Im}\,}
\newcommand{\real}{{\rm Re}\,}
\newcommand{\ds}{\displaystyle}
\newcommand{\epm}[2]{
 \raisebox{-0.5ex}{\shortstack[l]{$\scriptstyle+#1$\\$\scriptstyle-#2$}}
                    }

\newcommand{\prl}{Phys.~Rev. Lett.}
\newcommand{\prd}{Phys.~Rev.~D}
\newcommand{\plb}{Phys.~Lett.~B}
\newcommand{\npb}{Nucl.~Phys.~B}
\newcommand{\zpc}{Z.~Phys.~C}

\newlength{\miniwidth}
\newlength{\miniwidthplot}
\newlength{\nseparation}
\newenvironment{nfigure}[1]
        {\begin{figure}[#1]\hrule\vspace{\nseparation}\par}
        {\vspace{\nseparation}\par \hrule \end{figure}}
\newenvironment{ntable}[1]
        {\begin{table}[#1]\hrule\vspace{\nseparation}\par}
        {\vspace{\nseparation}\par \hrule \end{table}}

\begin{document}
~\\[-10truemm]
MPI-Ph/97-043  \hfill DESY 97-208 \\
TUM-HEP-281/97 \hfill hep-ph/9802202 \\
~\vspace{1truecm}
\begin{center}
{\LARGE Determination of the CKM angle $\gamma$ and $|V_{ub}/V_{cb}|$
from inclusive direct CP asymmetries and
branching ratios in charmless $B$ decays\footnote{Work supported by BMBF 
     under contract no.~06-TM-874.}}\\[2\baselineskip]
\textsl{Alexander Lenz\footnote{e-mail:alenz@MPPMU.MPG.DE},\\
Max-Planck-Institut f\"ur Physik --- Werner-Heisenberg-Institut,\\
F\"ohringer Ring 6, D-80805 M\"unchen, Germany, \\[2mm]  
Ulrich Nierste\footnote{e-mail:nierste@mail.desy.de},\\
DESY - Theory group, Notkestrasse 85, D-22607 Hamburg, Germany,\\[2mm]
and \\[2mm]
Gaby
Ostermaier\footnote{e-mail:Gaby.Ostermaier@feynman.t30.physik.tu-muenchen.de}\\
Physik-Department, TU M\"unchen, D-85747 Garching, Germany}
\end{center}
\vfill
\begin{center}
\textbf{\large Abstract} 
\end{center}
We have calculated inclusive direct CP-asymmetries for charmless
$B^{\pm}$--decays. After summing large logarithms to all orders the
CP-asymmetries in $\Delta S=0$ and $\Delta S=1$ decays are found as
\begin{eqnarray} 
    a_{CP}\lt( \Delta S=0 \rt) \; = \; \lt( 2.0 \epm{1.2}{1.0} \rt) \%,
      && \qquad 
    a_{CP}\lt( \Delta S=1 \rt) \; = \; \lt( -1.0 \pm 0.5 \rt) \% . \no
\end{eqnarray} 
These results are much larger than previous estimates based on a work
without summation of large logarithms. We further show that the
dominant contribution to $ a_{CP}\lt( \Delta S=0 \rt)$ is proportional
to $\sin \gamma \cdot |V_{cb}/V_{ub}|$.  The constraints on the apex
$(\ov{\rho},\ov{\eta})$ of the unitarity triangle obtained from these
two CP-asymmetries define circles in the
$(\ov{\rho},\ov{\eta})$-plane.  We have likewise analyzed the
information on the unitarity triangle obtainable from a measurement of
the average non-leptonic branching ratios $\ov{Br}(\Delta S=0)$,
$\ov{Br}(\Delta S=1)$ and their sum $\ov{Br}_{NL} (B \rightarrow
\textit{no charm})$. These CP-conserving quantities define circles
centered on the $\ov{\rho}$-axis of the $(\ov{\rho},\ov{\eta})$-plane.
We expect a determination of $|V_{ub}/V_{cb}|$ from $\ov{Br}_{NL} (B
\rightarrow\textit{no charm})$ to be promising.  Our results contain
some new QCD corrections enhancing $\ov{Br}(\Delta S=1)$, which now
exceeds $\ov{Br}(\Delta S=0)$ by roughly a factor of two.
\thispagestyle{empty}
\newpage
\section{Introduction}
CP-violation is a litmus test for the Standard Model, which
parametrizes all CP-violating quantities by a single parameter, the
complex phase in the Cabibbo-Kobayashi-Maskawa (CKM) matrix.  The
related amplitudes are further suppressed due to the smallness of CKM
elements and loop graphs, so that new physics effects may become
detectable.  CP-violating observables are commonly expressed in terms
of the angles $\alpha$, $\beta$ and $\gamma$ of the unitarity
triangle. Yet we can determine its shape not only from its angles, but
also from the length of its sides, which are obtained from
measurements of CP-conserving quantities.  This interplay is a special
feature of the CKM mechanism.  In order to overconstrain the unitarity
triangle one must find sufficiently many theoretically clean
observables.  While, for example, $\beta$ can be extracted without
hadronic uncertainties from the mixing-induced CP asymmetry in $B_d
\rightarrow J/\psi K_S$, the angle $\gamma$ is notoriously hard to
measure in experiments with $B_d$ and $B^{\pm}$ mesons.

Direct CP-violation in exclusive $B^{\pm}$-decays does not help to
determine any of the angles because of the unknown strong phases in
the decay amplitudes.  On the contrary direct \emph{inclusive}
CP-asymmetries can be cleanly predicted, because quark-hadron duality
allows the reliable calculation of strong interaction effects within
perturbation theory.  Such direct inclusive asymmetries have been
analyzed in \cite{bss,gh,sew,f,w,u} and mixing-induced inclusive
CP-asymmetries studied in \cite{bbd} are now investigated by the SLD
collaboration \cite{sld}. Semi-inclusive direct CP-asymmetries have
been studied in \cite{br}.  While inclusive final states are
experimentally difficult to identify, inclusive branching ratios are
huge compared to exclusive ones. As we will see in the following,
inclusive CP-asymmetries in charmless decays have a promising size, so
that it is worthwile to study them experimentally.  Further they can
be obtained from branching ratios only and therefore do not require an
asymmetric $B$-factory.

In this paper we calculate direct inclusive CP-asymmetries in
charmless $B^{\pm}$-decays extending our recent calculation of decay
rates in \cite{lno}. In \cite{lno} the corresponding branching ratios
have been calculated in renormalization group improved perturbation
theory including the dominant contributions of the next-to-leading
order. In the following section we set up our notations and summarize
previous work on the subject. In sect.~\ref{sect:s0} we analyze
$\Delta S=0$ decays. We discuss the relation of the CP-asymmetries to
the angles of the unitarity triangle and their impact on the
determination of the improved Wolfenstein parameters $\ov{\rho}$ and
$\ov{\eta}$. Here we also investigate the constraint on
$(\ov{\rho},\ov{\eta})$ imposed by a measurement of the average
branching ratio of $B$ and $\ov{B}$ decays with $\Delta S=0$.  In
sect.~\ref{sect:s1} we repeat the procedure for $\Delta S=1$ decays.
Readers mainly interested in phenomenology may draw their attention to
sect.~\ref{sect:phen}, where we will give numerical predictions for
the newly calculated quantities. In sect.~\ref{sect:phen} we also
predict the total charmless non-leptonic branching ratio of the $B$
meson and exemplify how the unitarity triangle is constructed from the
branching ratios and CP-asymmetries. Finally we summarize our
findings. An appendix contains details of our analytical results.

\section{Preliminaries}\label{sect:pr}
We start our discussion with $B$ decays corresponding to the quark
level transition $b \rightarrow q \ov{q} d$, $q=u,d,s,c$.  They are
triggered by the $|\Delta B|=1$, $|\Delta S|=0$ hamiltonian $H$:
\begin{eqnarray}
H &=&  \frac{G_F}{\sqrt{2}} 
        \left\{ 
        \sum_{j=1}^2 C_j \left( \xi_c^* Q_j^c + 
                                \xi_u^* Q_j^u \right) 
        - \xi_t^*    
          \sum_{j \in \mathcal{P}} C_j Q_j 
        \right\} + h.c. \; , \qquad \qquad
\xi_{q} \; = \; V_{q b}^* V_{q d}
\; .
\label{hd}
\end{eqnarray} 
Here $Q_{1,2}^{c,u}$ are the familiar current-current operators, which
originate from the tree-level $W$-exchange in $b \rightarrow c \ov{c}
d$ and $b \rightarrow u \ov{u} d$. Further $\mathcal{P}= \{ 3,\ldots
6, 8 \}$, and $Q_{3-6}$ and $Q_8$ are the penguin operators. More
details can be found in \cite{lno}, where the numerical values for the
Wilson coefficients $C_i$ are tabulated.  For the following we only
have to keep in mind that the coefficients $C_{3-6}$ and $C_{8}$
accompanying $\xi_t^*$ are much smaller in magnitude than $C_1$ and
$C_2$.

Now we express the decay rate for $b \rightarrow q \ov{q} d$ as 
\begin{eqnarray}
\Gamma &=& \frac{G_F^2 m_b^5}{64 \pi^3} \cdot \real \lt[ 
 \lt| \xi_u \rt|^2  \Gamma_{uu} + 
 \lt| \xi_t \rt|^2  \Gamma_{tt} +  
 \lt| \xi_c \rt|^2  \Gamma_{cc} +
 \xi_u \xi_c^* \Gamma_{uc} +
 \xi_t \xi_u^* \Gamma_{tu} +
 \xi_t \xi_c^* \Gamma_{tc} \rt] . \label{dec}
\end{eqnarray} 
The coefficients $\Gamma_{ij}$ encode the various contributions of the
different operators in \eq{hd}. For example in $b\rightarrow u \ov{u}
d$ the interference of the tree diagram of $Q_{2}^u$ in \fig{fig:tree}
with the penguin diagram of $Q_2^c$ with $q^\prime =c$ in
\fig{fig:peng} contributes to $\Gamma_{uc}$.  The average branching
ratio for the decay of $B^{\pm}$ into some inclusive final state $X$
reads
\begin{eqnarray}
\ov{Br} &=& \frac{  \Gamma \lt( B^+ \rightarrow X \rt) + 
 \Gamma \lt( B^- \rightarrow \ov{X} \rt) }{2 \Gamma_{tot}} . \label{defbr}
\end{eqnarray}
Similarly we define the CP-asymmetries as
\begin{eqnarray}
A_{CP} &=& \frac{1}{2} \lt[ Br \lt( B^+ \rightarrow X \rt) - 
 Br \lt( B^- \rightarrow \ov{X} \rt) \rt], \qquad \quad
a_{CP} \; = \; \frac{A_{CP}}{\ov{Br} } . \label{defacp}
\end{eqnarray}
Of course the average branching ratio in \eq{defbr} may also be
considered for $B_d$ and $\ov{B}_d$ instead of $B^+$ and $B^-$.  We do
not consider small spectator effects in this work, so that all given
formulae for $\ov{Br}$ likewise apply to the neutral $B$ mesons. We
will classify the inclusive final state $X$ by its strangeness quantum
number $S$. Hence if also $B_s$ mesons are included in the
consideration of $\ov{Br}$, the strangeness of $X$ must be corrected
for the non-zero strangeness of the spectator quark.

Our strategy is to express $\ov{Br}$ and the CP-asymmetries in
\eq{defacp} in terms of the $\Gamma_{ij}$'s. The constraints for the
CKM matrix obtained from measurements of $\ov{Br}$ and $a_{CP}$ are
most conveniently expressed in terms of the improved Wolfenstein
parameters $(\ov{\rho},\ov{\eta})$ \cite{blo}. The calculation of
$(\ov{\rho},\ov{\eta})$ from $\ov{Br}$ and $a_{CP}$ involves certain
combinations of the $\Gamma_{ij}$'s, for which we will derive compact
approximate formulae. The exact expressions for the $\Gamma_{ij}$'s
can be found in the appendix.

\begin{nfigure}{tb}
\begin{minipage}[t]{0.45\textwidth}
\centerline{\epsfxsize=0.4\textwidth \rotate[r]{\epsffile{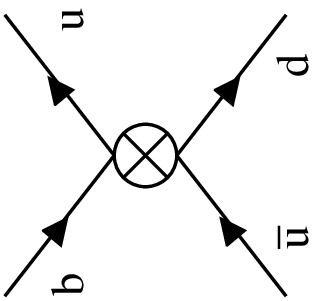}}}
\caption{Tree diagram of $Q^u_{1,2}$. }
\label{fig:tree}
\end{minipage}\hspace{2ex}
\begin{minipage}[t]{0.45\textwidth}
\centerline{\epsfysize=0.4\textwidth \rotate[r]{\epsffile{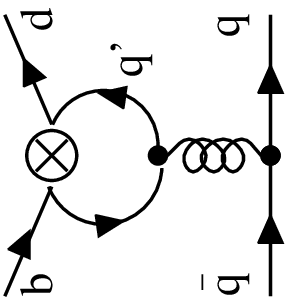}}}
\caption{Penguin diagram involving $Q_2^{q^\prime}$. It contributes the 
  absorptive part necessary for a non-zero $a_{CP}$.}
\label{fig:peng}
\end{minipage}
\end{nfigure}  

The three angles of the unitarity triangle are 
\begin{eqnarray}
\arg \lt( - \xi_t \xi_u^*  \rt) &=& \alpha, \quad \qquad
\arg \lt( - \xi_c \xi_t^*  \rt) \;=\; \beta, \quad  \qquad
\arg \lt( - \xi_u \xi_c^*  \rt) \;=\; \gamma .  \label{ckmangle}
\end{eqnarray}
Then $\ov{Br}$ and $A_{CP}$ are easily found as 
\begin{eqnarray}
\ov{Br} &=& F \cdot \frac{1}{|V_{cb}|^2} \cdot \lt[ 
     \lt| \xi_u \rt|^2  \Gamma_{uu} + 
     \lt| \xi_t \rt|^2  \Gamma_{tt} +  
     \lt| \xi_c \rt|^2  \Gamma_{cc} \rt. \nn 
&& \phantom{F \cdot \frac{1}{|V_{cb}|^2} \cdot} 
   - \lt.
     \lt| \xi_u \xi_c \rt|\, \cos \gamma \, \real \Gamma_{uc} -
     \lt| \xi_t \xi_u \rt|\, \cos \alpha \, \real \Gamma_{tu} -
     \lt| \xi_t \xi_c \rt|\, \cos \beta \, \real \Gamma_{tc} \rt] 
     \label{B} \\
A_{CP} &=& F \cdot \frac{1}{|V_{cb}|^2} \cdot \lt[ 
     -  \lt| \xi_u \xi_c \rt|\, \sin \gamma \, \imag \Gamma_{uc} -
     \lt| \xi_t \xi_u \rt|\, \sin \alpha \, \imag \Gamma_{tu} +
     \lt| \xi_t \xi_c \rt| \, \sin \beta \, \imag \Gamma_{tc} \rt] .
     \label{Acp}
\end{eqnarray}
The common factor $F$ reads
\begin{eqnarray}
F&=& \frac{B_{SL}^{exp}}{0.1045} \cdot 
      \lt[ 0.715 + 3.0 (x_c-0.3) + 11 (x_c-0.3)^2 
         \rt] \cdot 
      \lt[ 1 - 0.04 \log \frac{\mu}{m_b} \rt] . \label{deff}
\end{eqnarray}
Here $x_c=m_c/m_b$ and $\mu=O(m_b)$ is the renormalization scale.  $F$
is inverse proportional to the total decay rate $\Gamma_{tot}$, which
we calculate via $\Gamma_{tot}=\Gamma_{SL}/B_{SL}$ from the measured
semileptonic branching ratio $B_{SL}$.  The numerical approximation in
\eq{deff} holds to an accuracy of 1 \% in the range $0.25 \leq x_c
\leq 0.35$ and for variations of the renormalization scale $\mu$ in
the range $m_b/2 \leq \mu \leq 2 m_b $. The exact expression can be
found in the appendix.

From \eq{Acp} one can nicely verify that one needs two different CKM
structures and a non-zero absorptive part $\imag \Gamma_{ij}$ in order
to obtain a non-vanishing $A_{CP}$. It is known for long that the CPT
theorem correlates the CP-asymmetries for different subsets of final
states $X$ in \eq{defacp} \cite{gh,w}. For example $A_{CP} (\Delta
|C|=2, \Delta S=0)=-A_{CP} (\Delta |C|=0, \Delta S=0)$, where ($\Delta
|C|=2$, $\Delta S=0$) denotes the decay into the inclusive final state
with total strangeness zero containing a $c$ and a $\ov{c}$ quark,
while $\Delta |C|=0$ corresponds to a charmless final state. In the
following we will focus on charmless final state and omit ``$\Delta
|C|=0$'' in our notation. The non-zero contributions to $A_{CP}
(\Delta S=0)$ come from the absorptive parts of penguin diagrams (see
\fig{fig:peng}) involving the annihilation process
$(c,\ov{c})\rightarrow (q,\ov{q})$, $q=u,d,s$. We have illustrated the
leading $O(\alpha_s)$ contribution to $\imag \Gamma_{uc}$, $\imag
\Gamma_{tc}$ and $\imag \Gamma_{tu}$ in Figs.~\ref{fig:uc} and
\ref{fig:tctu}.  The results of all possible operator insertions into
Figs.~\ref{fig:uc} and \ref{fig:tctu} can be expressed in terms of a
single function $g(m_{q^\prime}/m_b, \mu/m_b)$, e.g.\ $\imag
\Gamma_{tc} \propto \imag g(x_c, \mu/m_b)$ (for details see the
appendix and \cite{lno}).  We will need some special values:
\begin{eqnarray}
g (0, 1) &=& -0.67 - 0.93 i, \qquad
g (x_c=0.3, 1) \;=\; -0.69 - 0.23 i, \qquad
g ( 1,1) \; = \; 0.28 \label{gnum}.
\end{eqnarray}
The imaginary part of $g$ is $\mu$-independent. Incidentally we will 
omit the second argument of $g$. 

\begin{nfigure}{tb}
\centerline{\epsffile{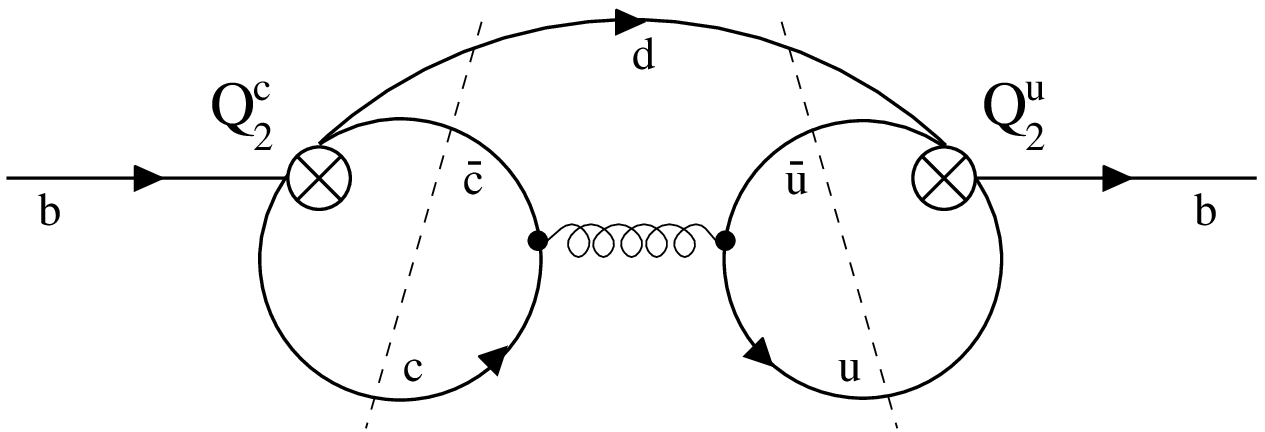} }
\caption{Diagram constituing $\imag \Gamma_{uc}$ to order $\alpha_s$
for $\Delta S=0$ decays. The right cut marks the final state
$u\ov{u}d$. The left cut denotes the absorptive part of the penguin 
diagram of \fig{fig:peng}.  }\label{fig:uc}
\end{nfigure}  
\begin{nfigure}{tb}
\centerline{\epsffile{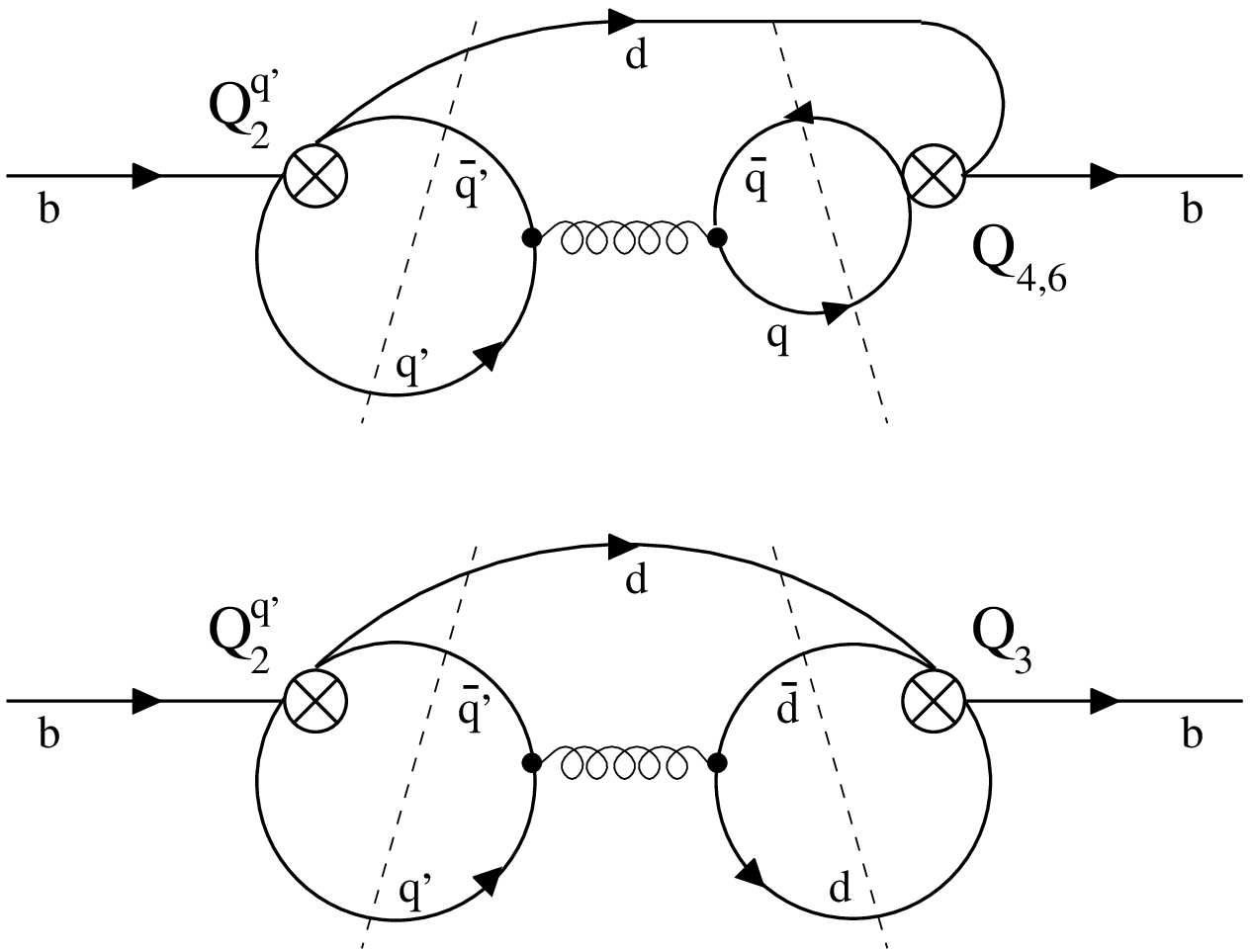} }
\caption{The diagrams show the contributions to $\imag \Gamma_{tc}$
and $\imag \Gamma_{tu}$ to order $\alpha_s$. $\imag \Gamma_{tc}$ 
corresponds to $q^\prime=c$ with the right cut denoting the final state
with $q=u,d$ or $s$. Likewise $\imag \Gamma_{tu}$ is obtained for 
$q^\prime=u$, but now also the left cut marks a possible charmless
final state. Then the right cut denotes the absorptive part of a
penguin diagram with internal quarks $q=u,d,s$ or $c$. The
contributions of this figure are suppressed with respect to those of 
\fig{fig:uc}, because they involve a small penguin coefficient $C_{3-6}$. 
}\label{fig:tctu}
\end{nfigure}

Let us now look at the CP-asymmetry related to a specific quark final
state, for definiteness we consider $u\ov{u}d$: The contribution from
$\Gamma_{uc}$ depicted in \fig{fig:uc} involves $Q_2^c$ and $Q_2^u$
and is therefore proportional to $C_2^2$, while $\Gamma_{tc}$ and
$\Gamma_{tu}$ involve $\xi_t^*$ and a small penguin coefficient
$C_{3-6}$ thereby.  Now $\imag \Gamma_{uc} \propto \imag g
(x_c)=-0.23$ for $x_c=0.3$.  The smallness of $\imag g (0.3)$ compared
to $\imag g (0)$ in \eq{gnum} reflects the fact that $\imag g (x_c) $
vanishes for $x_c \geq 1/2$.  Yet G\'erard and Hou \cite{gh} have made
the important observation that this kinematic suppression is absent in
the higher order contributions to $\Gamma_{uc}$, so that the result of
\fig{fig:uc} receives a correction of order $\alpha_s(m_b)/\pi \cdot
\imag g (0)/\imag g (x_c) \approx 30\%$. But these unsuppressed terms
cancel in the sum $A_{CP}\lt( u\ov{u}d\rt)+ A_{CP}\lt(s\ov{s}d\rt)+
A_{CP}\lt(d\ov{d}d\rt)=A_{CP} \lt(\Delta S=0\rt)=-A_{CP} \lt(\Delta
|C|=2, \Delta S=0\rt)$, because the latter asymmetry vanishes in the
kinematically forbidden region $x_c\geq 1/2$ \cite{w,gh}.  In this
work we will only calculate the inclusive CP asymmetries for charmless
$\Delta S=0$ and $\Delta S=1$ decays and therefore do not need to
include terms of order $\alpha_s^2$. This, however, is not true for
the separate inclusive CP-asymmetries $A_{CP}\lt(s\ov{s}d^\prime\rt)$
and $A_{CP}\lt(d\ov{d}d^\prime\rt)$, $d^\prime=d,s$, calculated to
order $C_2 C_{3-6} \alpha_s$ in \cite{f}. In addition the
$O(\alpha_s^2)$-contributions to these quantities involve the large
results of ``double penguin'' diagrams proportional to $C_2^2$
corresponding to the square of the diagram in \fig{fig:peng}.

Still there is an important difference between our calculations and
those in \cite{gh}: We use the effective hamiltonian of \eq{hd}, while
G\'erard and Hou perform their calculation in the full theory and
thereby invoke large logarithms, which are summed to all orders in our
approach. These large logarithms lead to an apparently large
contribution of order $\alpha_s^2$ in \cite{gh}, which had been found
to cancel the leading contribution of order $\alpha_s$ numerically, so
that the authors of \cite{gh} have claimed the total inclusive
asymmetries to be vanishingly small, of order of a few permille. As we
will see in the following, the correct summation of the large
logarithms leads to a different result: \emph{The inclusive
  CP-asymmetries $a_{CP} (\Delta S=0)$ and $a_{CP} (\Delta S=1)$ are
  sizeable, of the order of two and one percent, respectively.}

CP-asymmetries with resummed large logarithms have also been
calculated in \cite{u}, but for the case of a light $m_t \simeq
15~\mbox{GeV}$. In \cite{u} therefore no penguin operators $Q_{3-6}$
and $Q_8$ appear. The actual numerical results for $m_t \simeq
170~\mbox{GeV}$ are substantially different. In \cite{u} also the 
observation has been made that corrections of order $\alpha_s^2$ are
small for $a_{CP} (\Delta S=0)$ and $a_{CP} (\Delta S=1)$.
 
\section{$\mathbf{\Delta S=0}$ decays}\label{sect:s0}
We first look at the dominant contributions to $\ov{Br}$ and $a_{CP}$:
Keeping only the lowest nonvanishing order in $\alpha_s$ and
neglecting the contributions of the small penguin coefficients one
finds 
\begin{eqnarray}
\hspace{-1ex}
\ov{Br} \lt( \Delta S=0   \rt) &=& F \, |V_{ud}|^2 \, 
                \lt| \frac{V_{ub}}{V_{cb}} \rt|^2, \qquad
a_{CP} \lt( \Delta S=0 \rt) \;=\; - \imag \Gamma_{uc} \,
       \lt| \frac{V_{cd}}{V_{ud}} \rt| \, 
       \lt| \frac{V_{cb}}{V_{ub}} \rt| 
\, \sin \gamma.  \label{nopi}
\end{eqnarray}
Hence from $\ov{Br}$ one can determine $\lt| V_{ub}/V_{cb}\rt|$,
because $F$ and $V_{ud}$ are well-known. Likewise $a_{CP}$ measures
the product of $\sin \gamma $ and $\lt| V_{cb}/V_{ub} \rt|$. The
corrections to \eq{nopi} stemming from the penguin coefficients and
higher order corrections to $\ov{Br}$ are reliably calculable and small.
The best way to exploit \eq{nopi} and to include these corrections is
the use of the improved Wolfenstein parameters $A$, $\lambda$,
$\ov{\rho}$ and $\ov{\eta}$ \cite{blo}.  Then $\ov{Br}$ of \eq{B}
reads
\begin{eqnarray}
\ov{Br} \lt( \Delta S=0 \rt)  &=& L_B 
         \lt[ (\ov{\rho} -\ov{\rho}_0)^2 + \ov{\eta}^2 - K \rt]. 
\label{brw} 
\end{eqnarray}
Here 
\begin{subequations}
\label{re} 
\begin{eqnarray}
L_B &=&  F \, \lambda^2 \, \lt[ \Gamma_{tt} + \Gamma_{uu} 
           - \real \Gamma_{tu}  \rt] \label{re1} \\
\ov{\rho}_0 &=& \frac{2\Gamma_{tt} + 
      \real \lt[ \Gamma_{uc} - \Gamma_{tc} -\Gamma_{tu}   \rt] }{ 
        2 \lt[ \Gamma_{tt}+\Gamma_{uu} - \real \; \Gamma_{tu}   \rt]}
           \label{re2} \\
K &=& \frac{1}{4 \, \lt[ \Gamma_{tt} + \Gamma_{uu} - \real \Gamma_{tu}
             \rt]^2 }\cdot  \lt[ 
      - 4 \Gamma_{uu} \Gamma_{tt} +  4 \Gamma_{uu} \real \Gamma_{tc} 
      + \lt[ \real \Gamma_{uc} \rt]^2   - 
        2 \real \Gamma_{uc} \, \real \Gamma_{tu} \rt. \nn
&& \phantom{\frac{1}{4 \, \lt[ \Gamma_{tt} + \Gamma_{uu} - \real \Gamma_{tu}
             \rt]}   }
    - 4 \Gamma_{cc} \Gamma_{uu} +  4 \Gamma_{cc} \real \Gamma_{tu} 
      + \lt[ \real \Gamma_{tu} \rt]^2   - 
        2 \real \Gamma_{tu} \, \real \Gamma_{tc}  \nn
&& \phantom{ \frac{1}{4 \, \lt[ \Gamma_{tt} + \Gamma_{uu} - \real \Gamma_{tu}
             \rt] }   }
\lt. 
      - 4 \Gamma_{tt} \Gamma_{cc} +  4 \Gamma_{tt} \real \Gamma_{uc} 
      + \lt[ \real \Gamma_{tc} \rt]^2   - 
        2 \real \Gamma_{tc} \, \real \Gamma_{uc} \rt] . 
\end{eqnarray}
\end{subequations}
We stress here that our notation of $\ov{Br}(\Delta S=0)$ only
comprises non-leptonic decays, but not the semileptonic decay
$B\rightarrow X_u \ell \ov{\nu}_{\ell}$, which is measured in a
different way. In addition to the quark final states $u\ov{u}d$,
$s\ov{s}d$ and $d\ov{d}d$ we have also included the decay $b
\rightarrow d\,g$, which gives a small contribution of order 3\% to
$\ov{Br}(\Delta S=0)$, but has a non-negligible impact on $K$ and
$\rho_0$.  Notice that the Wolfenstein parameter $A$ drops out in
\eq{re}. The corrections to the formulae in \eq{re} are of order
$\lambda^6$ and therefore negligible. From \eq{brw} one sees that the
measurement of the CP-conserving quantity $\ov{Br}$ defines a circle
in the $(\ov{\rho},\ov{\eta})$-plane centered at $(\ov{\rho}_0,0)$
with radius $R_{B}$, where
\begin{eqnarray}
R_{B}^2 &=& \frac{\ov{Br} \lt( \Delta S=0 \rt)}{L_B} + K . \label{rad}
\end{eqnarray}
The center $(\ov{\rho}_0,0)$ and $K$ are independent of the measured
$\ov{Br}$, they vanish in the limit considered in \eq{nopi}. For the
constraint from the CP asymmetry we likewise define 
\begin{eqnarray}
L_a &=& \frac{ \imag \lt[ \Gamma_{tc} - \Gamma_{tu} -\Gamma_{uc}
  \rt]}{ \Gamma_{tt} + \Gamma_{uu} - \real \Gamma_{tu} } .\label{la}
\end{eqnarray}
Then 
\begin{eqnarray}
A_{CP} \lt( \Delta S=0 \rt) &=& L_a L_B  \, \ov{\eta}, \qquad \qquad
a_{CP} \lt( \Delta S=0 \rt) \; = \; L_a \, 
  \frac{\ov{\eta} }{ ( \ov{\rho} -\ov{\rho}_0 )^2 + \ov{\eta}^2 - K }
\label{acpk} .
\end{eqnarray}
Again the corrections to \eq{acpk} are suppressed with four powers of
$\lambda$ and therefore negligible. Now \eq{acpk} reveals that a
measurement of $a_{CP}$ likewise fixes a circle in the
$(\ov{\rho},\ov{\eta})$-plane.  This new circle is centered at
$(\ov{\rho}_0,\ov{\eta}_0)$ and its radius equals $R_a$ with
\begin{eqnarray}
\ov{\eta}_0 &=& \frac{L_a}{2 a_{CP} \lt( \Delta S=0 \rt) }, \qquad \qquad
R_a^2 \;=\; \ov{\eta}_0^2 + K. \label{circacp}   
\end{eqnarray}
Again in the approximation with $K=\ov{\rho}_0=0$ adopted in \eq{nopi} the
circle defined by \eq{acpk} is centered exactly on the
$\ov{\eta}$-axis.  Its radius equals $\ov{\eta}_0$, so that it passes
through the origin.  In \eq{nopi} $\sin \gamma$ comes with $\lt|
V_{cb} / V_{ub} \rt|$, which is inverse proportional to
$\sqrt{\ov{\rho}^2+\ov{\eta}^2}$.  The geometrical construction of
$\gamma$ from $a_{CP}$ corresponding to \eq{nopi} is therefore done by
intersecting the circle in \eq{acpk} with the one centered at $(0,0)$
stemming from any measurement of $|V_{ub}/V_{cb}|$. Of course any
other information on the apex $(\ov{\rho},\ov{\eta})$ of the unitarity triangle
can be included in the usual way, and ideally the hyperbola from
$\epsilon_K$ \cite{hn,bjw}, the circle from $\Delta m_B$ \cite{bjw}
and the new circles in \eq{brw} and \eq{acpk} intersect in the same
point $(\ov{\rho},\ov{\eta})$ --- or we may find new physics.

We close this section by giving compact approximate expressions for
the quantities in \eq{re} and \eq{la}, which enter the circles
defined by \eq{re2}, \eq{rad} and \eq{circacp}:
\begin{eqnarray}
&&\hspace{-5ex}
\begin{array}{|rcr@{\:}l@{\:}l|l|}
\hline &&&&&\\[-2mm]
\ds L_B & = & \ds \big( 0.0362 & 
               \ds + \: 0.151 \, (x_c-0.3)   & 
               \ds + \: 0.58 \,  (x_c-0.3)^2 \big) & 
\\[1mm]
           && \ds \cdot \big( 1      &  
              \ds - \: 0.12 \, \ln \frac{\mu}{m_b} & 
              \ds + \: 0.02 \, \ln^2 \frac{\mu}{m_b} \big)
              & \ds \textrm{uncertainty:\ }  0.80 \% 
\\[3mm]\hline &&&&&\\[-2mm]
\ds L_a & = & \ds \big(0.00734 & \ds
                  - \: 0.0905 \, (x_c-0.3) & \ds 
                  + \: 0.220 \,  (x_c-0.3)^2 \big) & 
\\[1mm] 
           && \ds \cdot \big( 1 & \ds 
                  - \: 0.22 \, \ln \frac{\mu}{m_b} & \ds 
                  + \: 0.08 \, \ln^2 \frac{\mu}{m_b} \big)
              & \ds \textrm{uncertainty:\ } 3.6 \% 
\\[3mm] \hline &&&&&\\[-2mm] 
\ds K   & = & \ds \big( -0.0133 & \ds 
                  + \: 0.0219 \, (x_c-0.3) & \ds 
                  + \: 0.032 \, (x_c-0.3)^2 \big) & 
\\[1mm] 
           && \ds \cdot \big( 1 & \ds 
                  - \: 0.44 \, \ln \frac{\mu}{m_b} & \ds 
                  + \: 0.16 \, \ln^2 \frac{\mu}{m_b} \big)
              & \ds \textrm{uncertainty:\ } 3.3 \% 
\\[3mm] \hline &&&&&\\[-2mm]
\ov{\rho}_0 & = & \big( -0.0254 & \ds 
                  + \: 0.034 \, (x_c-0.3) & \ds 
                  + \: 0.12 \, (x_c-0.3)^2 \big)  &
\\[1mm] 
           &&& \hfill  \ds \cdot \big( 1  
                 & \ds 
                 + \: 0.03 \, \ln^2 \frac{\mu}{m_b} \big)
              & \ds \textrm{uncertainty:\ } 1.8 \% 
\\[3mm]\hline
\end{array}\label{num}
\end{eqnarray}
In the last column we have listed the error of our approximate
formulae compared to the exact expressions for the range $0.25 \leq
x_c \leq 0.35$ and $0.5\leq \mu/m_b \leq 2.0$. Further
$\alpha_s(M_Z)=0.118$ and $L_B$ is calculated with $B_{SL}=0.1045$.
The $\mu$-dependence in \eq{num} results from the truncation of the
perturbation series and is small in $L_B$, for which the dominant
next-to-leading order corrections are known. A future calculation of
the full $O(\alpha_s)$ corrections to $\ov{Br}$ and the
$O(\alpha_s^2)$ corrections to $A_{CP}$ will change the numbers in the
first brackets in \eq{num} by a term of order $\alpha_s(m_b)/\pi$ and
will reduce the size of the coefficients of $\ln(\mu/m_b)$ and $\ln^2
(\mu/m_b)$.

\section{$\mathbf{\Delta S=1}$ decays}\label{sect:s1}
To obtain the  $|\Delta S|=1$ hamiltonian from \eq{hd} we must simply
replace $\xi_q$ by $\xi_q^{(s)} = V_{q b}^* V_{q s}$. Instead of 
\eq{ckmangle} we invoke the CKM angles
\begin{eqnarray}
\arg \lt( - \xi_t^{(s)} \xi_u^{(s)*}  \rt) & = & - \phi \; = \; 
               -   \lt( \gamma - \lambda^2 \, \ov{\eta} \rt) 
               + O \lt( \lambda^4 \rt) , \nn
\arg \lt( - \xi_c^{(s)} \xi_t^{(s)*}  \rt) & = & - \varepsilon \; = \; 
             - \lambda^2 \, \ov{\eta} \lt( 1 + \lambda^2 (1- \ov{\rho})  \rt) 
               + O \lt( \lambda^6 \rt) , \nn 
\arg \lt( - \xi_u^{(s)} \xi_c^{(s)*}  \rt) & = & \gamma - \pi  
                     + O \lt( \lambda^4 \rt) . \no
\end{eqnarray}
Hence the corresponding unitarity triangle with angles $\pi-\gamma$,
$\phi$ and $\varepsilon$ is squashed. In the limit of vanishing
penguin coefficients one has 
\begin{eqnarray}
A_{CP} \lt( \Delta S=1 \rt) & \propto & |V_{ub} V_{cb}| \sin \gamma . \no
\end{eqnarray}
Yet an approximate formula for $a_{CP}\lt( \Delta S=1 \rt)$ similar to
\eq{nopi} cannot be found, because the tree-level contribution to $
\ov{Br} \lt( \Delta S=1 \rt)$ is CKM suppressed and the different
$\Gamma_{ij}$'s are equally important. An analogue of \eq{nopi} would
involve more than one CKM angle. 

Next we express $\ov{Br}$, $A_{CP}$ and $a_{CP}$ as in \eq{brw} and
\eq{acpk}:
\begin{eqnarray}
\ov{Br} \lt( \Delta S=1 \rt)  &=& L_B^\prime 
         \lt[ (\ov{\rho} -\ov{\rho}_0^\prime)^2 + \ov{\eta}^2 - K^\prime \rt],
\label{brw1}  \\
A_{CP} \lt( \Delta S=1 \rt) &=& L_a^\prime L_B^\prime  
        \, \ov{\eta}, \qquad \qquad
a_{CP} \lt( \Delta S=1 \rt) \; = \; L_a^\prime \, 
  \frac{\ov{\eta} }{ ( \ov{\rho} -\ov{\rho}_0^\prime )^2 + \ov{\eta}^2
    - K^\prime }
\label{acpk1} .
\end{eqnarray}
The primed coefficients read
\begin{eqnarray}
L_B^\prime &=& F\, \lambda^4 \lt(1+\lambda^2\rt) 
    \lt( \Gamma_{uu} + \Gamma_{tt} - 
               \real \Gamma_{tu} \rt) , \nn
\ov{\rho}_0^\prime &=& 
  \frac{- 2 \Gamma_{tt} + 
        \real \lt( \Gamma_{tu} + \Gamma_{tc} - \Gamma_{uc} \rt)}{ 
        2\, \lambda^2 \lt(1+\lambda^2\rt) 
              \lt( \Gamma_{uu} + \Gamma_{tt} - \real \Gamma_{tu}  \rt)
              }, \nn 
L_a^\prime &=& 
   \frac{ \imag \lt( \Gamma_{uc}+ \Gamma_{tu} - \Gamma_{tc}  \rt)}{
        \lambda^2 \lt( 1+ \lambda^2 \rt) 
        \lt( \Gamma_{uu} + \Gamma_{tt} - \real \Gamma_{tu}  \rt)} , \nn
K^\prime &=& \frac{ -\lt( 1-\lambda^2 \rt) \lt( 
      \Gamma_{cc} + \Gamma_{tt} - 
       \real \Gamma_{tc} \rt) }{\lambda^4 \lt( 1+\lambda^2 \rt) 
    \lt( \Gamma_{uu} + \Gamma_{tt} - \real \Gamma_{tu} \rt) }
    + 
     \frac{ \lt[ - 2 \Gamma_{tt} + 
          \real \lt( \Gamma_{tu} + \Gamma_{tc} - \Gamma_{uc} 
                \rt) \rt]^2 
          }{4\, \lambda^4 \lt( 1+\lambda^2 \rt)^2 
    \lt( \Gamma_{uu} + \Gamma_{tt} - \real \Gamma_{tu} \rt)^2 }
 .\label{lrk}
\end{eqnarray}
In $L_B^\prime $, $\ov{\rho}_0^\prime $ and $K^\prime $ we have kept
corrections of order $\lambda^2$ and omitted corrections of order
$\lambda^4$ and higher in accordance with the adopted improved
Wolfenstein approximation \cite{blo}. The powers of $\lambda$ in the
denominators of $\ov{\rho}_0^\prime$ and $K^\prime$ are partially
numerically compensated by the smallness of the penguin coefficients
entering the $\Gamma_{ij}$'s in the numerators.
The corresponding approximate formulae read
\begin{eqnarray}
&&\hspace{-5ex}
\begin{array}{|rcr@{\:}l@{\:}l|l|}
\hline &&&&&\\[-2mm]
L_B^\prime & = & \ds \big(0.00185 & 
         \ds + \: 0.0077
\, (x_c-0.3) &
         \ds + \: 0.03 \,  (x_c-0.3)^2 \big) & 
\\[1mm]
         && \ds \cdot \big(
         1 & \ds - \: 0.12 \, \ln \frac{\mu}{m_b} &
         \ds + \:0.02 \, \ln^2 \frac{\mu}{m_b} \big)
              & \ds \textrm{uncertainty:\ } 1.0  \% 
\\[3mm]\hline &&&&&\\[-2mm]
L_a^\prime & = & \ds \big(-0.144 & 
         \ds + \: 1.77
\, (x_c-0.3) &
         \ds - \: 4  \,  (x_c-0.3)^2 \big) & 
\\[1mm]
         && \ds \cdot \big(
         1 & \ds - \: 0.22 \, \ln \frac{\mu}{m_b} &
         \ds + \:0.08 \, \ln^2 \frac{\mu}{m_b} \big)
              & \ds \textrm{uncertainty:\ } 4.3  \% 
\\[3mm]\hline &&&&&\\[-2mm]
K^\prime & = & \ds \big(-5.08 & 
         \ds + \: 8.4 
\, (x_c-0.3) &
         \ds + \: 12 \,  (x_c-0.3)^2 \big) & 
\\[1mm]
         && \ds \cdot \big(
         1 & \ds - \: 0.44 \, \ln \frac{\mu}{m_b} &
         \ds + \: 0.16 \, \ln^2 \frac{\mu}{m_b} \big)
              & \ds \textrm{uncertainty:\ } 2.9 \% 
\\[3mm]\hline &&&&&\\[-2mm]
\ov{\rho}_0^\prime & = & \ds \big(0.498 & 
         \ds - \: 0.67
\, (x_c-0.3) &
         \ds - \: 2.4 \,  (x_c-0.3)^2 \big) & 
\\[1mm]
         &&& \ds \hfill  \cdot \big(
         1 & \ds + \:0.03 \, \ln^2 \frac{\mu}{m_b} \big) 
              & \ds \textrm{uncertainty:\ } 1.8 \% 
\\[3mm]\hline
\end{array}\label{num1}
\end{eqnarray}
Here we emphasize that in \eq{num1} we have not only included the
final states with quark contents $u\ov{u}s$, $d\ov{d}s$ and
$s\ov{s}s$, but also the decay $b \rightarrow s\,g$, which gives a
non-negligible contribution to $\ov{Br}( \Delta S=1)$ in \eq{brw1}.
Further we had to include the contributions to the decay rate stemming
from the square of the penguin diagram in \fig{fig:peng}.  These
contributions are of order $\alpha_s^2$, but are proportional to
$C_2^2$ and the fourth power of $\lambda$. They belong to
$\Gamma_{cc}$ in \eq{dec} and amount to 13 \% of $\ov{Br}( \Delta
S=1)$. The large contributions of penguin operators and penguin
diagrams imply that $\ov{Br}( \Delta S=1)$ is quite insensitive to 
$\ov{\rho}$ and $\ov{\eta}$. This is reflected by the large value of
$K^\prime$ in \eq{num1}. Consequently $\ov{Br}( \Delta S=1)$ becomes only 
a useful observable to constrain $(\ov{\rho},\ov{\eta})$ once its 
experimental accuracy is better than 10 \%. 

The geometrical constructions of the circles obtained from $\ov{Br}(
\Delta S=1)$ and $a_{CP} (\Delta S=1)$ is done in a completely
analogous way to sect.~\ref{sect:s0}. One merely has to replace the
unprimed quantities in \eq{rad} and \eq{circacp} by the primed ones of
\eq{num1} to obtain the $\Delta S=1$ parameters $R_B^\prime$,
$\ov{\eta}_0^\prime$ and $R_a^\prime$. Since the denominator of
$a_{CP} (\Delta S=1)$ in \eq{acpk} depends very weakly on $\ov{\rho}$
and $\ov{\eta}$, $a_{CP} (\Delta S=1)$ is almost proportional to
$\ov{\eta}$ and both radius $R_a^\prime$ and offset
$\ov{\eta}_0^\prime$ of the corresponding circle are very large. This
is very different from the situation in $\Delta S=0$ decays.
 
Finally we mention that 
\begin{eqnarray}
A_{CP} \lt( \Delta S=1 \rt) &=& -A_{CP} \lt( \Delta S=0 \rt)  
\label{ckmcp}
\end{eqnarray}
for $m_s=0$. This is a consequence of the CKM mechanism of CP violation.   
The relation in \eq{ckmcp} receives corrections by terms of order 
$m_s^2/m_b^2$ and $m_s/m_b \cdot \alpha_s(m_b)/\pi$. A larger
deviation from \eq{ckmcp} would be an experimental sign of 
non-standard CP-violation outside the quark mass matrix.

\section{Phenomenology}\label{sect:phen}
In this section we give numerical predictions for the branching ratios
and CP asymmetries and exemplify, how the apex $(\ov{\rho},\ov{\eta})$
is constructed from future measurements of $\ov{Br}_{NL} (B \rightarrow
\textit{no charm})$, $a_{CP}(\Delta S=0)$ and $a_{CP}(\Delta S=1)$.

First we express $\ov{Br}_{NL} (B \rightarrow \textit{no charm})$
analogously to \eq{brw} and \eq{brw1}:
\begin{eqnarray}
\ov{Br}_{NL} \lt( B \rightarrow \textit{no charm} \rt) &=& 
 \ov{Br} \lt( \Delta S=0 \rt) + \ov{Br} \lt( \Delta S=1 \rt) 
 \; = \; \widetilde{L}_B \lt( \ov{\rho}^2 + \ov{\eta}^2 - \widetilde{K} \rt) 
.\label{brw2}
\end{eqnarray}
There is no dependence on $\gamma$ here, i.e.\ $\widetilde{\rho}_0=0$,
for the same reason as \eq{ckmcp}.  It is easy to relate
$\widetilde{L}_B$ and $\widetilde{K}$ to $L_B$, $L_B^\prime$, $K$ and
$K^\prime$. The approximate formulae read
\begin{eqnarray}
&&\hspace{-5ex}
\begin{array}{|rcr@{\:}l@{\:}l|l|}
\hline &&&&&\\[-2mm]
\widetilde{L}_B  & = & \ds \big(0.0380 & 
         \ds + \: 0.158
\, (x_c-0.3) &
         \ds + \: 0.6 \,  (x_c-0.3)^2 \big) & 
\\[1mm]
         && \ds \cdot \big(
         1 & \ds - \: 0.12 \, \ln \frac{\mu}{m_b} &
         \ds + \:0.02 \, \ln^2 \frac{\mu}{m_b} \big)&
         \ds \textrm{uncertainty:\ } 0.80 \% 
\\[3mm]\hline &&&&&\\[-2mm]
\widetilde{K} & = & \ds \big(-0.272 & 
         \ds + \: 0.46
\, (x_c-0.3) &
         \ds + \: 0.7 \,  (x_c-0.3)^2 \big) & 
\\[1mm]
         && \ds \cdot \big(
         1 & \ds - \: 0.42 \, \ln \frac{\mu}{m_b} &
         \ds + \: 0.15  \, \ln^2 \frac{\mu}{m_b} \big)&
         \ds \textrm{uncertainty:\ } 2.7 \% . 
\\[3mm]\hline
\end{array}\label{num2}
\end{eqnarray}
As usual the last column lists the error of the approximate formulae  
for the range $0.25 \leq x_c \leq 0.35$ and $0.5\leq \mu/m_b \leq 2.0$
with $\alpha_s(M_Z)=0.118$ and $B_{SL}=0.1045$. 

\subsection{Numerical predicitions}
Next we predict the average branching ratios and the CP asymmetries as
a function of $|V_{ub}/V_{cb}|$ and $\gamma$. For this we recall the
relation of these quantities to the improved Wolfenstein parameters
$(\ov{\rho},\ov{\eta})$ \cite{blo}:
\begin{eqnarray} 
\lt| \frac{V_{ub}}{V_{cb}}\rt| &=& \lt( 1+\frac{\lambda^2}{2} \rt) \,
                        \lambda \, \sqrt{\ov{\rho}^2 + \ov{\eta}^2  },
                        \qquad  \quad
\tan \gamma \; = \; \frac{\ov{\eta}}{\ov{\rho}} .
\end{eqnarray} 
The predictions for the branching ratios can be found in
Tabs.~\ref{tab:noc}, \ref{tab:bs0} and \ref{tab:bs1}. Then $a_{CP}$ is
tabulated in Tabs.~\ref{tab:as0} and \ref{tab:as1}. The range of
$\gamma$ in the tables is the one favoured by the standard
next-to-leading order \cite{hn,bjw} analysis of the unitarity triangle
from $\epsilon_K$ and $\Delta m_B$.  The central values in the tables
correspond to the following set of input parameters:
\begin{eqnarray}
&& \hspace{-1ex}
\begin{array}[b]{lll}
& \ds
x_c = 0.29 ,
& \ds 
\mu = m_b = 4.8~\mbox{GeV} ,  \\[1mm]
\ds 
\alpha_s \lt( M_Z \rt) = 0.118 , 
& \ds 
m_t \lt( m_t \rt) = 168~\mbox{GeV}, 
& \ds 
B_{SL}^{exp} =  0.1045 . 
\end{array}
\label{inp}
\end{eqnarray}
Here $m_b$ is the one-loop pole mass. The errors in the tables
correspond to a variation of $x_c=m_c/m_b$ and the renormalization
scale $\mu$ within the range 
\begin{eqnarray}
0.25 \leq x_c \leq 0.33, && \qquad \qquad  0.5\leq \mu/m_b \leq 2.0 .\no 
\end{eqnarray}
The corresponding error bars are added in quadrature. The experimental
uncertainty in $\alpha_s$ has a smaller impact on the listed quantities,
the errors  of the remaining input quantities in \eq{inp} have a
negligible influence.

\begin{ntable}{tb}
\begin{tabular}{l|l|l|l|l}
$\ds \lt| \frac{V_{ub}}{V_{cb}} \rt|=0.06$ & 
$\ds \lt| \frac{V_{ub}}{V_{cb}} \rt|=0.07$ & 
$\ds \lt| \frac{V_{ub}}{V_{cb}} \rt|=0.08$ & 
$\ds \lt| \frac{V_{ub}}{V_{cb}} \rt|=0.09$ & 
$\ds \lt| \frac{V_{ub}}{V_{cb}} \rt|=0.10$ 
 \\[4mm] \hline \hline &&&& \\[-2mm]
  0.0127\epm{0.0057}{0.0034}
& 0.0136\epm{0.0058}{0.0035}
& 0.0147\epm{0.0060}{0.0036}
& 0.0159\epm{0.0062}{0.0038}
& 0.0172\epm{0.0064}{0.0040}
\end{tabular}
\caption{The total nonleptonic charmless branching ratio 
$\ov{Br}_{NL} (B \rightarrow \textit{no charm})$ as a function of 
$|V_{ub}/V_{cb}|$. It is independent of $\gamma$.
}\label{tab:noc}
\end{ntable}

\begin{ntable}{tb}
\begin{tabular}{@{\hspace{1.6pt}}r||l|l|l|l|l}
&
$\ds \lt| \frac{V_{ub}}{V_{cb}} \rt|=0.06$ & 
$\ds \lt| \frac{V_{ub}}{V_{cb}} \rt|=0.07$ & 
$\ds \lt| \frac{V_{ub}}{V_{cb}} \rt|=0.08$ & 
$\ds \lt| \frac{V_{ub}}{V_{cb}} \rt|=0.09$ & 
$\ds \lt| \frac{V_{ub}}{V_{cb}} \rt|=0.10$ 
 \\[4mm] \hline \hline &&&&& \\[-2mm]
$\ds \gamma= 60^\circ $ 
& 0.0032\epm{0.0007}{0.0005}
& 0.0041\epm{0.0009}{0.0007}
& 0.0052\epm{0.0011}{0.0008}
& 0.0064\epm{0.0014}{0.0010}
& 0.0077\epm{0.0017}{0.0012}
\\[1mm]  \hline &&&&& \\[-2mm]
$\ds \gamma= 75^\circ $ 
& 0.0031\epm{0.0007}{0.0005}
& 0.0040\epm{0.0009}{0.0007}
& 0.0050\epm{0.0011}{0.0008}
& 0.0062\epm{0.0014}{0.0010}
& 0.0075\epm{0.0016}{0.0012}
\\[1mm]  \hline &&&&& \\[-2mm]
$\ds \gamma= 90^\circ $ 
& 0.0029\epm{0.0007}{0.0005}
& 0.0038\epm{0.0009}{0.0006}
& 0.0048\epm{0.0011}{0.0008}
& 0.0060\epm{0.0013}{0.0010}
& 0.0073\epm{0.0016}{0.0012}
\\[1mm]  \hline &&&&& \\[-2mm]
$\ds \gamma= 105^\circ $ 
& 0.0028\epm{0.0007}{0.0005}
& 0.0037\epm{0.0009}{0.0006}
& 0.0047\epm{0.0011}{0.0008}
& 0.0058\epm{0.0013}{0.0010}
& 0.0071\epm{0.0016}{0.0012}
\\[1mm]  \hline &&&&& \\[-2mm]
$\ds \gamma= 120^\circ $ 
& 0.0027\epm{0.0007}{0.0005}
& 0.0035\epm{0.0008}{0.0006}
& 0.0045\epm{0.0010}{0.0008}
& 0.0056\epm{0.0013}{0.0009}
& 0.0069\epm{0.0015}{0.0011}
\end{tabular}
\caption{The average inclusive branching ratio into nonleptonic final states 
with zero strangeness,  $\ov{Br} \lt( \Delta S=0 \rt)$, as a function 
of $|V_{ub}/V_{cb}|$ and $\gamma$.
}\label{tab:bs0}
\end{ntable}

\begin{ntable}{tb}
\begin{tabular}{@{\hspace{1.6pt}}r||l|l|l|l|l}
&
$\ds \lt| \frac{V_{ub}}{V_{cb}} \rt|=0.06$ & 
$\ds \lt| \frac{V_{ub}}{V_{cb}} \rt|=0.07$ & 
$\ds \lt| \frac{V_{ub}}{V_{cb}} \rt|=0.08$ & 
$\ds \lt| \frac{V_{ub}}{V_{cb}} \rt|=0.09$ & 
$\ds \lt| \frac{V_{ub}}{V_{cb}} \rt|=0.10$ 
 \\[4mm] \hline \hline &&&&& \\[-2mm]
$\ds \gamma= 60^\circ $ 
& 0.0095\epm{0.0051}{0.0029}
& 0.0095\epm{0.0051}{0.0029}
& 0.0095\epm{0.0051}{0.0029}
& 0.0095\epm{0.0051}{0.0029}
& 0.0096\epm{0.0051}{0.0029}
\\[1mm]  \hline &&&&& \\[-2mm]
$\ds \gamma= 75^\circ $ 
& 0.0096\epm{0.0051}{0.0029}
& 0.0096\epm{0.0051}{0.0029}
& 0.0097\epm{0.0051}{0.0029}
& 0.0097\epm{0.0051}{0.0030}
& 0.0097\epm{0.0051}{0.0030}
\\[1mm]  \hline &&&&& \\[-2mm]
$\ds \gamma= 90^\circ $ 
& 0.0097\epm{0.0051}{0.0030}
& 0.0098\epm{0.0051}{0.0030}
& 0.0098\epm{0.0051}{0.0030}
& 0.0099\epm{0.0051}{0.0030}
& 0.0099\epm{0.0051}{0.0030}
\\[1mm]  \hline &&&&& \\[-2mm]
$\ds \gamma= 105^\circ $ 
& 0.0098\epm{0.0051}{0.0030}
& 0.0099\epm{0.0051}{0.0030}
& 0.0100\epm{0.0051}{0.0030}
& 0.0101\epm{0.0052}{0.0030}
& 0.0102\epm{0.0052}{0.0030}
\\[1mm]  \hline &&&&& \\[-2mm]
$\ds \gamma= 120^\circ $ 
& 0.0100\epm{0.0051}{0.0030}
& 0.0100\epm{0.0052}{0.0030}
& 0.0101\epm{0.0052}{0.0030}
& 0.0102\epm{0.0052}{0.0030}
& 0.0103\epm{0.0052}{0.0030}
\end{tabular}
\caption{The average inclusive branching ratio into nonleptonic final states 
with strangeness one,  $\ov{Br} \lt( \Delta S=1 \rt)$, as a function 
of $|V_{ub}/V_{cb}|$ and $\gamma$.}\label{tab:bs1}
\end{ntable}

\begin{ntable}{tb}
\begin{tabular}{r||l|l|l|l|l}
&
$\ds \lt| \frac{V_{ub}}{V_{cb}} \rt|=0.06$ & 
$\ds \lt| \frac{V_{ub}}{V_{cb}} \rt|=0.07$ & 
$\ds \lt| \frac{V_{ub}}{V_{cb}} \rt|=0.08$ & 
$\ds \lt| \frac{V_{ub}}{V_{cb}} \rt|=0.09$ & 
$\ds \lt| \frac{V_{ub}}{V_{cb}} \rt|=0.10$ 
 \\[4mm] \hline \hline &&&&& \\[-2mm]
$\ds \gamma= 60^\circ $ 
& 0.021\epm{0.010}{0.009}
& 0.019\epm{0.009}{0.008}
& 0.017\epm{0.009}{0.007}
& 0.016\epm{0.008}{0.007}
& 0.014\epm{0.007}{0.006}
\\[1mm]  \hline &&&&& \\[-2mm]
$\ds \gamma= 75^\circ $ 
& 0.024\epm{0.012}{0.010}
& 0.022\epm{0.011}{0.009}
& 0.020\epm{0.010}{0.008}
& 0.018\epm{0.009}{0.008}
& 0.016\epm{0.008}{0.007}
\\[1mm]  \hline &&&&& \\[-2mm]
$\ds \gamma= 90^\circ $ 
& 0.026\epm{0.013}{0.011}
& 0.023\epm{0.012}{0.010}
& 0.021\epm{0.011}{0.009}
& 0.019\epm{0.010}{0.008}
& 0.017\epm{0.009}{0.007}
\\[1mm]  \hline &&&&& \\[-2mm]
$\ds \gamma= 105^\circ $ 
& 0.026\epm{0.013}{0.011}
& 0.023\epm{0.012}{0.010}
& 0.021\epm{0.011}{0.009}
& 0.019\epm{0.010}{0.008}
& 0.017\epm{0.009}{0.007}
\\[1mm]  \hline &&&&& \\[-2mm]
$\ds \gamma= 120^\circ $ 
& 0.024\epm{0.012}{0.010}
& 0.022\epm{0.011}{0.009}
& 0.019\epm{0.010}{0.008}
& 0.018\epm{0.009}{0.007}
& 0.016\epm{0.008}{0.007}
\end{tabular}
\caption{The inclusive indirect CP-asymmetry for charmless final states 
with zero strangeness, $a_{CP}(\Delta S=0)$. 
}\label{tab:as0}
\end{ntable}

\begin{ntable}{tb}
\begin{tabular}{r||l|l|l|l|l}
&
$\ds \lt| \frac{V_{ub}}{V_{cb}} \rt|=0.06$ & 
$\ds \lt| \frac{V_{ub}}{V_{cb}} \rt|=0.07$ & 
$\ds \lt| \frac{V_{ub}}{V_{cb}} \rt|=0.08$ & 
$\ds \lt| \frac{V_{ub}}{V_{cb}} \rt|=0.09$ & 
$\ds \lt| \frac{V_{ub}}{V_{cb}} \rt|=0.10$ 
 \\[4mm] \hline \hline &&&&& \\[-2mm]
$\ds \gamma= 60^\circ $ 
& -0.007\epm{0.003}{0.003}
& -0.008\epm{0.003}{0.004}
& -0.009\epm{0.004}{0.004}
& -0.010\epm{0.004}{0.005}
& -0.012\epm{0.005}{0.005}
\\[1mm]  \hline &&&&& \\[-2mm]
$\ds \gamma= 75^\circ $ 
& -0.008\epm{0.003}{0.003}
& -0.009\epm{0.004}{0.004}
& -0.010\epm{0.004}{0.004}
& -0.011\epm{0.004}{0.005}
& -0.013\epm{0.005}{0.005}
\\[1mm]  \hline &&&&& \\[-2mm]
$\ds \gamma= 90^\circ $ 
& -0.008\epm{0.003}{0.003}
& -0.009\epm{0.004}{0.004}
& -0.010\epm{0.004}{0.004}
& -0.012\epm{0.005}{0.005}
& -0.013\epm{0.005}{0.006}
\\[1mm]  \hline &&&&& \\[-2mm]
$\ds \gamma= 105^\circ $ 
& -0.007\epm{0.003}{0.003}
& -0.009\epm{0.003}{0.004}
& -0.010\epm{0.004}{0.004}
& -0.011\epm{0.004}{0.005}
& -0.012\epm{0.005}{0.005}
\\[1mm]  \hline &&&&& \\[-2mm]
$\ds \gamma= 120^\circ $ 
& -0.007\epm{0.003}{0.003}
& -0.008\epm{0.003}{0.003}
& -0.009\epm{0.003}{0.004}
& -0.010\epm{0.004}{0.004}
& -0.011\epm{0.004}{0.005}
\end{tabular}
\caption{The inclusive indirect CP-asymmetry for charmless final states 
with strangeness one, $a_{CP}(\Delta S=1)$.}\label{tab:as1}
\end{ntable}

From a comparison of \tab{tab:bs1} with \tab{tab:bs0} one realizes
that charmless non-leptonic $B$--decays occur preferably with
$\Delta S=1$, with $\ov{Br} \lt( \Delta S=1 \rt)$ exceeding $\ov{Br} \lt(
\Delta S=0 \rt)$ by roughly a factor of two:
\begin{eqnarray}
\frac{ \ov{Br} \lt( \Delta S=0 \rt) }{\ov{Br} \lt( \Delta S=1 \rt)}
&=& 0.50 \pm 0.12 \qquad \mbox{for~~} \frac{|V_{ub}|}{|V_{cb}|}=0.08. 
\no
\end{eqnarray}
Most of the dependence on $x_c$ stems from the normalization factor
$F$ and cancels in ratios of different $\ov{Br}$'s.  The
$\mu$-dependence of $\ov{Br} \lt( \Delta S=1 \rt)$ is much larger than
the one of $\ov{Br} \lt( \Delta S=0 \rt)$ leading to larger error bars
in \tab{tab:bs0}.  This comes from the penguin dominance of
$\ov{Br} \lt( \Delta S=1 \rt)$ and the fact that current-current type
radiative corrections to penguin operators have not been calculated
yet. The newly calculated contributions enhance $\ov{Br} \lt( \Delta
S=1 \rt)$ explaining the increase of $\ov{Br}_{NL} (B \rightarrow
\textit{no charm})$ in \tab{tab:noc} compared to the result in
\cite{lno}. In order to obtain the total charmless branching ratio
$\ov{Br} (B \rightarrow \textit{no charm})$ one must add twice the
charmless semileptonic branching ratio $Br (B\rightarrow X_u \ell
\ov{\nu}_{\ell})$, for $\ell=e$ and $\ell=\mu$ \cite{nir}:
\begin{eqnarray}
Br \lt( B\rightarrow X_u \ell \ov{\nu}_{\ell} \rt) &=& 
\lt( 0.0012\epm{0.0002}{0.0002} \rt) 
 \cdot \lt( \frac{ \lt| V_{ub}/V_{cb} \rt| }{0.08} \rt) ^2
   \no . 
\end{eqnarray}
Hence for the input of \eq{inp} one finds
from \tab{tab:noc}:
\begin{eqnarray}
\ov{Br} (B \rightarrow \textit{no charm}) &=& 
	0.0097 \epm{0.0051}{0.0030} +
  \lt( 0.0073 \epm{0.0012}{0.0009} \rt)
  \lt( \frac{\lt| V_{ub}/V_{cb} \rt|}{0.08} \rt)^2
   \no . 
\end{eqnarray}
The present experimental result for the total charmless branching ratio 
reads 
\begin{eqnarray}
\ov{Br}^{exp} (B \rightarrow \textit{no charm}) &=& 0.002 \pm 0.041,  \no
\end{eqnarray}
obtained in \cite{n} from CLEO data \cite{cleo}.  

We conclude that the measurement of $\ov{Br} \lt( \Delta S=0 \rt)$
provides a competitive method to determine $|V_{ub}/V_{cb}|$ compared
to the standard analysis from semileptonic decays. Once a complete
next-to-leading order calculation is done for the $\Delta S=1$ decays,
the error bars in \tab{tab:noc} will reduce significantly and
$\ov{Br}_{NL} (B \rightarrow \textit{no charm})$ will likewise become
a promising observable to measure $|V_{ub}/V_{cb}|$.

The most important results of our calculations, however, are those
listed in \tab{tab:as0} and \tab{tab:as1}. Adding the errors stemming 
from the uncertainties in $|V_{ub}/V_{cb}|$ and $\gamma$ in quadrature
to the ones already included in the tables, we predict:
\begin{eqnarray}
a_{CP}\lt( \Delta S=0 \rt) \; = \; \lt( 2.0 \epm{1.2}{1.0} \rt) \%,
&& \qquad 
a_{CP}\lt( \Delta S=1 \rt) \; = \; \lt( -1.0 \pm 0.5 \rt) \% .
\label{resacp}
\end{eqnarray}
These results have to be contrasted with those of Table~1 in
\cite{gh}, where predictions for the $a_{CP}$'s are given, which are
five times smaller than those in \eq{resacp}.  This discrepancy is
partly due to the fact that we sum large logs to all orders whereas
this has not been done in \cite{gh}.  It is further related to the use
of an extremely small $|\sin \gamma \cdot V_{cb}/V_{ub}|$ in
\cite{gh}.  The reduction of the $\mu$-dependence in \tab{tab:as0} and
\tab{tab:as1} requires the calculation of $\imag \Gamma$ to order
$\alpha_s^2$. The corresponding diagrams are obtained by dressing
\fig{fig:uc} and \fig{fig:tctu} with an extra gluon. A part of this
calculation has been performed in \cite{sew}.  In a perfect experiment
the detection of $a_{CP}\lt( \Delta S=0 \rt)=2\%$ with $\ov{Br} \lt(
\Delta S=0 \rt)= 5 \cdot 10^{-3}$ at the $3\sigma$ level requires the
production of $4.5 \cdot 10^6$ $B^{\pm}$ mesons. This should be worth
looking at by our experimental colleagues.  Finally we remark that our
results satisfy
\begin{eqnarray}
a_{CP} \lt( \Delta S=1 \rt) \cdot  \ov{Br} \lt( \Delta S=1 \rt) &=& 
- a_{CP} \lt( \Delta S=0 \rt) \cdot \ov{Br} \lt( \Delta S=0 \rt) . \no
\end{eqnarray}
as required by \eq{ckmcp}.

\subsection{Construction of $\mathbf{(\ov{\rho},\ov{\eta})}$}
In this section we exemplify how the circles in the
$(\ov{\rho},\ov{\eta})$-plane will be constructed from a future
measurement of $\ov{Br}_{NL} (B \rightarrow \textit{no charm})$,
$\ov{Br} \lt( \Delta S=0 \rt)$, $\ov{Br} \lt( \Delta S=1 \rt)$,
$a_{CP} \lt( \Delta S=0 \rt)$ and $a_{CP} \lt( \Delta S=1 \rt)$.

We first show this construction for the CP-conserving quantities.  We
assume that the three charmless non-leptonic branching ratios are
measured as
\begin{eqnarray} 
\ov{Br}_{NL} \lt(B \rightarrow \textit{no charm} \rt) &=& 1.47 \, \% ,\nn 
\ov{Br} \lt( \Delta S=0 \rt) \; = \; 0.50 \, \%, && \qquad \qquad   
\ov{Br} \lt( \Delta S=1 \rt) \; = \; 0.97 \, \% . \no 
\end{eqnarray} 
For illustration we assume an experimental error of $5\, \% $ in all
quantities and neglect the present theoretical uncertainty here by
setting $\mu=m_b$ and $x_c=0.29$. The three circles are defined by
\eq{brw}, \eq{brw1} and \eq{brw2}. To draw the circles we must only
read off the coefficients from \eq{num}, \eq{num1} or \eq{num2} and
calculate the radii $R_B$, $R_B^\prime$ and $\widetilde{R}_B$ from
\eq{rad}. The results are shown in \fig{fig:br}. 
\begin{nfigure}{tb}
\vspace{-2cm}
\centerline{\epsfysize=0.9\textwidth \epsffile{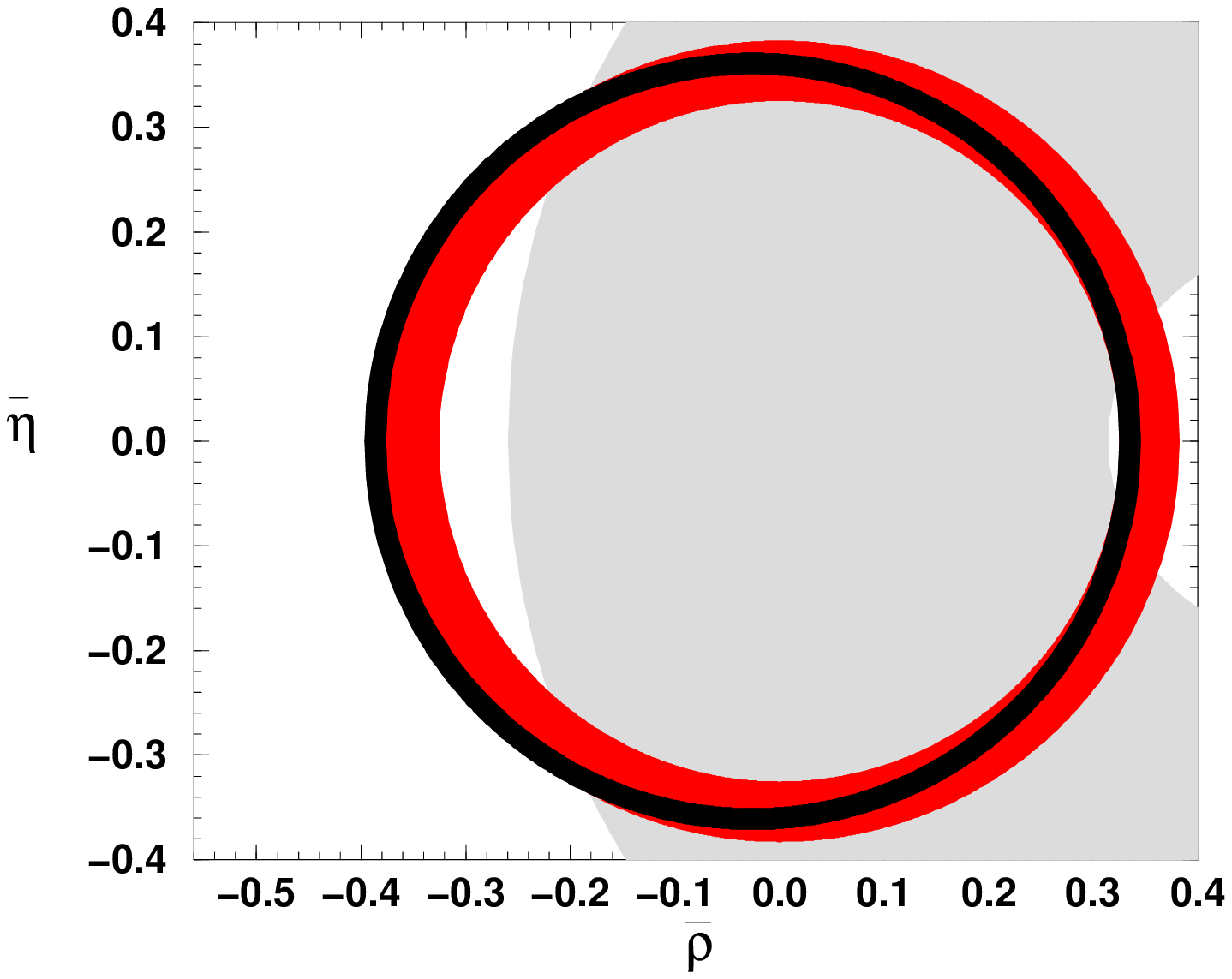}}
\caption{The black annulus shows the region allowed by 
  a measurement of $\ov{Br} \lt( \Delta S=0 \rt)$. The dark shading
  corresponds to $\ov{Br}_{NL} (B \rightarrow \textit{no charm})$ and
  the lightly shaded area shows the constraint on
  $(\ov{\rho},\ov{\eta})$ obtained from $\ov{Br} \lt( \Delta S=1
  \rt)$.
 }\label{fig:br}
\end{nfigure}  
The figure reveals that $\ov{Br} \lt( \Delta S=0 \rt)$ is a very good
observable for the phenomenology of the unitarity triangle. This
remains true even if the actual theoretical uncertainty of $20\, \%$
is included.  By contrast $\ov{Br} \lt( \Delta S=1 \rt)$ is not very
sensitive to $(\ov{\rho}, \ov{\eta})$ and thereby yields a much poorer
information on the unitarity triangle. Still the center
$(\ov{\rho}_0,0)$ of the circle largely deviates from the origin, so
that upper or lower bounds on $\ov{Br} \lt( \Delta S=1 \rt)$ could
help to exclude a part of the $(\ov{\rho},\ov{\eta})$-plane allowed by
other observables. Also \fig{fig:br} shows that $\ov{Br}_{NL} (B
\rightarrow \textit{no charm})$ is a very useful observable to
determine $\sqrt{\ov{\rho}^2+\ov{\eta}^2}$, once the large
$\mu$-dependence of the entries in \tab{tab:noc} is reduced by a
complete next-to-leading order calculation of the $\Delta S=1$ decay
rates.

The circles from the CP-asymmetries are likewise obtained from
\eq{circacp}. Here the measured value of $a_{CP}$ enters the
$\ov{\eta}$-coordinate $\ov{\eta}_0$ or $\ov{\eta}_0^\prime$ of the
center of the circle and its radius $R_a$ or $R_a^\prime$. For the
CP-asymmetries we assume an experimental precision of $20 \, \% $ and
\begin{eqnarray}
a_{CP} \lt( \Delta S=0 \rt) \; = \; 2.0 \, \% ,&& \qquad  \qquad  
a_{CP} \lt( \Delta S=1 \rt) \; = \; -1.0 \, \% . \no
\end{eqnarray}
\begin{nfigure}{tb}
\vspace{-2cm}
\centerline{\epsfysize=0.9\textwidth \epsffile{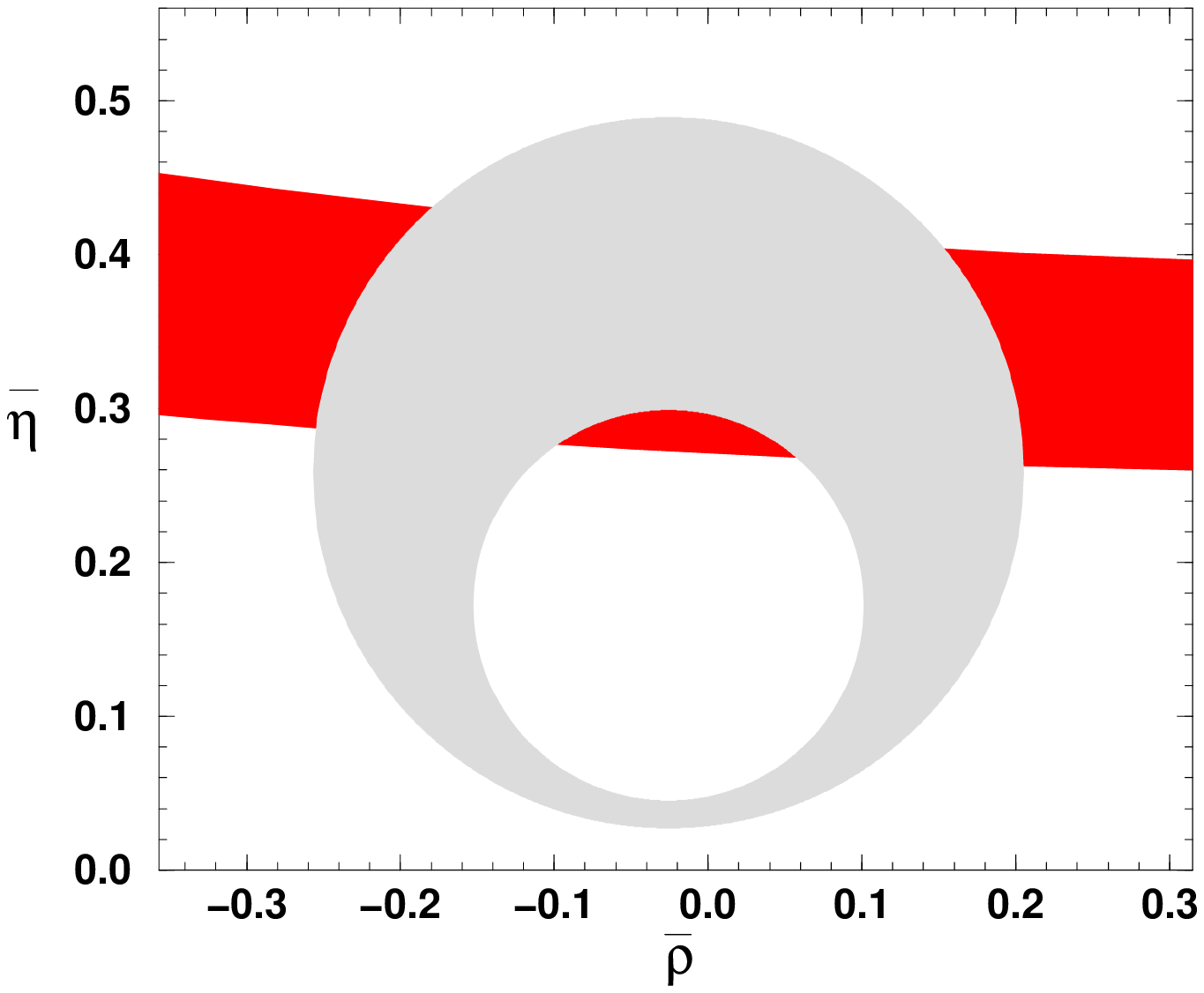}}
\caption{The lightly shaded area shows the constraint stemming 
from $a_{CP} \lt( \Delta S=0 \rt)$ and the dark shading marks the 
area allowed from $a_{CP} \lt( \Delta S=1 \rt)$.}\label{fig:acp}
\end{nfigure}  
The results are displayed in \fig{fig:acp}.  If one switches off the
effects of penguin operators, the circle from $a_{CP} \lt( \Delta S=0
\rt)$ touches the $\ov{\rho}$--axis in the point $(0,0)$. The distance
of the points on the circle to the origin is therefore proportional to
$\sin \gamma$, so that $a_{CP} \lt( \Delta S=0 \rt)$ measures $\sin
\gamma/\sqrt{\ov{\rho}^2+\ov{\eta}^2}$ in this limit as found in
\eq{nopi}. The circle from $a_{CP} \lt( \Delta S=1 \rt)$, however,
looks totally different: $\ov{\eta}_0$ and $R_a^\prime$ are so large
that only a small fraction of the circle can be seen in \fig{fig:acp}.
$a_{CP} \lt( \Delta S=1 \rt)$ weakly depends on $\ov{\rho}$ and yields
good information on $\ov{\eta}$. Hence from \fig{fig:acp} we learn
that inclusive CP-asymmetries yield interesting information on the
unitarity triangle, which is complementary to the one obtained from
other observables in the $B$ system. Alternatively one can multiply
$a_{CP}$ with the measured $\ov{Br}$ and obtain $A_{CP}$ of
\eq{defacp}, which defines a horizontal straight line in the
$(\ov{\rho},\ov{\eta})$-plane (see \eq{acpk} and \eq{acpk1}).

\section{Ten messages from this work}
\newcounter{mess}
\begin{list}{
             \raggedleft \arabic{mess})}
            {\usecounter{mess}
             \setlength{\rightmargin}{0pt}
            }
\item Inclusive direct CP-asymmetries in charmless $B^{\pm}$--decays are
  larger than previously believed:
  \begin{eqnarray} 
    a_{CP}\lt( \Delta S=0 \rt) \; = \; \lt( 2.0 \epm{1.2}{1.0} \rt) \%,
      && \quad 
    a_{CP}\lt( \Delta S=1 \rt) \; = \; \lt( -1.0 \pm 0.5 \rt) \% . \no
  \end{eqnarray} 
\item The dominant contribution to $a_{CP}\lt( \Delta S=0 \rt)$
  satisfies
  \begin{eqnarray} 
    a_{CP}\lt( \Delta S=0 \rt) &\propto& 
      \frac{\sin \gamma}{\lt|V_{ub}/V_{cb}\rt|} 
  \end{eqnarray} 
     with small and calculable corrections. 
\item The constraints on the apex $(\ov{\rho},\ov{\eta})$ of the
     unitarity triangle obtained from a measurement of $ a_{CP}\lt(
     \Delta S=0 \rt)$ and $ a_{CP}\lt( \Delta S=1 \rt)$ are circles in
     the $(\ov{\rho},\ov{\eta})$-plane. These constraints are
     complementary to the information from other observables in $K$
     and $B$ physics.
\item Inclusive direct CP-asymmetries are theoretically clean: The
  uncertainties can be controlled and systematically reduced by higher
  order calculations.
\item The CP-conserving observables $\ov{Br} \lt( \Delta S=0
      \rt)$, $\ov{Br} \lt( \Delta S=1 \rt)$ and $\ov{Br}_{NL} (B
      \rightarrow \textit{no charm})$ define circles in the 
      $(\ov{\rho},\ov{\eta})$-plane centered on the $\ov{\rho}$-axis.  
\item $\ov{Br} \lt( \Delta S=0 \rt)$ is well suited to determine 
      $|V_{ub}/V_{cb}|$, with little sensitivity to $\gamma$. 
\item ``Double penguin'' contributions, which are part of the 
      next-to-next-to-leading order, enhance 
      $\ov{Br} \lt( \Delta S=1 \rt)$ by $13\,\%$. 
\item The present incomplete next-to-leading order (NLO) result
      imposes a large $\mu$-de\-pen\-dence on $\ov{Br} \lt( \Delta S=1
      \rt)$. Here a calculation of all NLO corrections to penguin
      operator matrix elements is necessary.\label{pom} 
\item $\ov{Br} \lt( \Delta S=1 \rt)$ exceeds $\ov{Br} \lt( \Delta
      S=0 \rt)$ by roughly a factor of two. 
\item The determination of $|V_{ub}/V_{cb}|$ from 
      $\ov{Br}_{NL} (B \rightarrow \textit{no charm})$ is competitive 
      to the standard method from semileptonic decays, once the 
      NLO calculation mentioned in \ref{pom}) has been done.
\end{list}

\section*{Acknowledgements} 
A.L.~appreciates many stimulating discussions with Iris Abt. 
U.N.~thanks Andrzej Buras for his hospitality at the TUM, where
part of this work has been done. We thank him and Ahmed Ali for
proofreading the manuscript.

\appendix 

\section{Exact formulae}
Here we show how the quantities entering sect.~\ref{sect:s0} and
sect.~\ref{sect:s1} are related to the results of \cite{lno}. These
expression are useful for readers who are not satisfied with the
approximate formulae for $F$, $L_B$, $L_a$, etc.  They are also
helpful, if one wants to calculate the branching ratios and
CP-asymmetries in extensions of the Standard Model. Then one needs to
change the Wilson coefficients entering the $\Gamma_{ij}$'s
accordingly.

Now $F$ \cite{nir} reads 
\begin{eqnarray}
F &=& \frac{G_F^2 m_b^5 |V_{cb}|^2 }{64 \pi^3 \Gamma_{tot}} 
       \lt( 1 + \frac{\lambda_1}{2 m_b^2} \rt) 
\; = \; \frac{G_F^2 m_b^5 |V_{cb}|^2 }{64 \pi^3 } 
       \cdot \frac{B_{SL}^{exp}}{\Gamma_{SL}} \lt( 1 + 
       \frac{\lambda_1}{2 m_b^2} \rt)  \nn
  & = & \frac{3 B_{SL}^{exp}}{f_1 (x_c^2) 
          \lt[ 1 + \alpha_s (\mu) /(2 \pi) h_{SL} (x_c) \rt] 
          - 6\, (1-x_c^2)^4 \, \lambda_2/m_b^2   } \label{fex}
\end{eqnarray}
in terms of the notation of \cite{lno}.  Here we have used the common
trick to evaluate $\Gamma_{tot}=\Gamma_{SL}/B_{SL}$ via the
semileptonic rate and the experimental value of the semileptonic
branching ratio $B_{SL}$.  This eliminates various uncertainties
associated with the theoretical prediction for $\Gamma_{tot}$.  The
non-perturbative corrections involving the kinetic energy parameter
$\lambda_1$ has been factored out in \eq{fex}, because $\lambda_1$
cancels in $\ov{Br}$ and $A_{CP}$.

Likewise for the decay rates corresponding to the quark level
transition $b\rightarrow q \ov{q} d^\prime $, $q=u,d,s$ and
$d^\prime=d,s$, one has
\begin{eqnarray}
\Gamma_{uu} &=& t \sum_{i,j=1}^2 C_i C_j 
   \lt[ b_{ij} \lt( 1 - 6 \frac{\lambda_2}{m_b^2} \rt) + 
   \delta b_{ij}  
   + \frac{\alpha_s}{4 \pi} \, 2 \,  
               \real \lt[ h_{ij} + g_{ij} ( 0 ) \rt] \rt] \nn
\Gamma_{uc} &=& t \frac{\alpha_s}{2\pi} C_2^2 g_{22} (x_c) \nn
\Gamma_{tu} &=& - 2 \!\!\! \sum_{\scriptstyle i=1,2 \atop 
                           \scriptstyle j=3,\ldots 6} \!\! C_i C_j 
 \lt[ t \, b_{ij} \lt( 1 - 6 \frac{\lambda_2}{m_b^2} \rt) + 
      t \, \delta b_{ij} + \frac{\alpha_s}{4 \pi} \lt(  
                    g_{ij} ( 0 ) +t g_{ji}^* \lt( x_c \rt) \rt) \rt]  
                - C_8 C_2 \,  \frac{\alpha_s}{2\pi}  \,  b_{28} t \nn
\Gamma_{tc} &=& -2\!\!\!  \sum_{\scriptstyle i=1,2 \atop 
                          \scriptstyle j=3,\ldots 6} \!\! C_i C_j 
                \frac{\alpha_s}{4 \pi} g_{ij} (x_c) \nn 
\Gamma_{tt} &=&  \sum_{i,j=3,\ldots 6} \!\! C_i C_j \lt[
                 b_{ij} \lt( 1 - 6 \frac{\lambda_2}{m_b^2} \rt) + 
                 \delta b_{ij} \rt] + \frac{\alpha_s}{2 \pi} C_8  
                 \sum_{j=3,\ldots 6}  C_j b_{j8} \nn
\Gamma_{cc} &=& \lt( \frac{\alpha_s}{4 \pi} \rt)^2 C_2^2 
                \, k_{22} \lt( x_c,x_c, \frac{\mu}{m_b} \rt) .
    \label{capp}
\end{eqnarray}
Here $t=1$ for $b\rightarrow u \ov{u} d^\prime $, while $t=0$ for
$b\rightarrow s \ov{s} d^\prime $ and $b\rightarrow d \ov{d} d^\prime
$. The $C_j$'s, $\alpha_s$ and the loop functions $h_{ij},g_{ij}$ in
\eq{capp} are understood to be evaluated at the scale $\mu=O(m_b)$.
The $\Gamma_{ij}$'s depend sizeably on $\mu$ and $x_c$ as indicated in
the approximate formulae in sect.~\ref{sect:s0}. Further they depend
on $m_t$ and $M_W$, this dependence, however, is marginally small.
$g_{22} (x_c)=g (x_c,\mu/m_b)$ is the fundamental penguin functions 
entering all $g_{ij}$'s. 
For this work we have newly calculated 
\begin{eqnarray}
 g_{42} \lt(x_c, \frac{\mu}{m_b} \rt) & = &
   g_{62} \lt(x_c, \frac{\mu}{m_b} \rt)
   \;=\; n_f g \lt( 0 , \frac{\mu}{m_b} \rt) + 
               g \lt( x_c , \frac{\mu}{m_b} \rt) + 
               g \lt( 1 , \frac{\mu}{m_b} \rt) , 
           \qquad  n_f=3, \nn
g_{32} \lt(x, \frac{\mu}{m_b} \rt) & = &
   g \lt( 0 , \frac{\mu}{m_b} \rt) + 
   g \lt( 1 , \frac{\mu}{m_b} \rt)  , \qquad \quad 
g_{52} \lt(x, \frac{\mu}{m_b} \rt) \; = \; 0, 
\nn
  g \lt( 1, \frac{\mu}{m_b}  \rt) & = &
  - \frac{16}{27} \ln \frac{\mu}{m_b} + \frac{98}{8} - 
    \frac{8}{\sqrt{3}} \pi + \frac{32}{81} \pi^2 .
\label{newgs}
\end{eqnarray}
These quantities correspond to the diagrams of \fig{fig:tctu} with
$q^\prime=u$, $q=u,d,s,c,b$ and the left cut marking the final state
$u\ov{u}d$.  For the remaining $g_{ij}$'s we refer to \cite{lno},
where also analytic formulae for $g (x, \mu/m_b)$ and the $b_{ij}$'s
and $h_{ij}$'s \cite{acmp} can be found. In \eq{capp} the leading
nonperturbative corrections are also included, the $\delta b_{ij}$'s
\cite{bbsv} depend on $\lambda_2=0.12$ GeV${}^2$.  The values in
\eq{newgs} correspond to the NDR scheme, the vanishing of $g_{52}$
involves in addition the standard finite renormalization of $Q_5$
introduced in \cite{cfmrs} and related to the definition of the
``effective'' coefficient $C_8$.  

Another new result is $\Gamma_{cc}$ in \eq{capp}. We have calculated
the ``double penguin'' contribution stemming from the square of
\fig{fig:peng} with $q^\prime=c$. Although being of order $\alpha_s^2$
this term is numerically relevant in $\Delta S=1$ decays, because it
is proportional to $C_2^2$ and the tree-level result is CKM
suppressed.  We have also included $\Gamma_{cc}$ in the $\Delta S=0$
coefficients of \eq{num}. Approximately one finds
\begin{eqnarray}
\hspace{-10pt}
k_{22} \lt( x_c,x_c, x_{\mu}\rt) &=& \lt( 1- \frac{r}{6} \rt)
  \big[  1.52
         - \: 11.5 \, (x_c-0.3) 
          + \: 7 \,  (x_c-0.3)^2   \no
\\[1mm] 
&& \quad +   \lt( 1.84 - \: 5.6 \, (x_c-0.3) - 
       \: 19 \, (x_c-0.3)^2 \rt) 
 \ln x_{\mu} + 0.79 \ln^2 x_{\mu} \big] . \label{defk22}
\end{eqnarray}
with $r=0$ for the quark final states $u\ov{u}d$, $u\ov{u}s$,
$s\ov{s}d$, $d\ov{d}s$, and $r=1$ for $d\ov{d}d$ and $s\ov{s}s$.  The
result in \eq{defk22} receives corrections of order $(x_c-0.3)^3$
and reproduces $k_{22}$ with an error of 2.6 \% for $0.25 \leq x_c \leq
0.35$ and $0.5\leq \mu/m_b \leq 2.0$.

Our results for $\ov{Br} \lt( \Delta S=0 \rt)$ and $\ov{Br} \lt(
\Delta S=1 \rt)$ also include the decay rates for $b \rightarrow s\,
g$ and $b \rightarrow d\, g$. Here to order $\alpha_s$ all
$\Gamma_{ij}$'s are zero except for
\begin{eqnarray}
\Gamma_{tt} \lt( b \rightarrow s\,  g \rt) \;=\; 
\Gamma_{tt} \lt( b \rightarrow d\,  g \rt) 
&=&   \frac{8}{3} \frac{\alpha_s (\mu)}{\pi} C_8^2. \no
\end{eqnarray}

The approximate formulae in \eq{num}, \eq{num1} and \eq{num2} further
correspond to $\alpha_s (M_Z)=0.118$ corresponding to $\alpha_s
(\mu=4.8 \mathrm{GeV})=0.216$.  The dependence on $\alpha_s (M_Z)$ is
non-negligible, but smaller than the $\mu$-dependence.

When calculating $A_{CP}$ for the inclusive $\Delta S=0$ or $\Delta
S=1$ final state, we must add the $\Gamma_{ij}$'s for $b\rightarrow u
\ov{u} d^\prime $, $b\rightarrow s \ov{s} d^\prime $ and $b\rightarrow
d \ov{d} d^\prime $.  $\imag \Gamma_{tu}$ contains $\imag g(0)$,
which, however, cancels when summing $\imag \Gamma_{tu}$ for the three
decay modes $b \rightarrow u \ov{u} d^\prime$, $b \rightarrow s \ov{s}
d^\prime$ and $b \rightarrow d \ov{d} d^\prime$, so that $A_{CP}
(\Delta S=0)$ and $A_{CP} (\Delta S=1)$ vanish for $x_c \geq 1/2$ as
required by the CPT theorem. The cancellation takes place when summing
the contributions of different cuts of the diagrams in \fig{fig:tctu}
as found in \cite{gh}.

One comment is in order here: The terms of order $\alpha_s$ in
\eq{capp} depend on the renormalization scheme. This originates from
the fact that when renormalizing $H$ in \eq{hd} one already uses the
unitarity relation $\xi_u+ \xi_c+ \xi_t=0$. After using this relation
to eliminate, say, $\xi_c$ in \eq{dec} one finds the coefficients of
$|\xi_u|^2$, $|\xi_t|^2$ and $\xi_t \xi_u^*$ scheme independent.
Consequently by changing the scheme one can shift terms in $A_{CP}$ in
\eq{Acp} from e.g.\ the term proportional to $\sin \gamma$ to the one
multiplying $\sin \beta$. This scheme ambiguity, however, is
suppressed by a factor of $C_{3-6}/C_{1,2}$ with respect to the
dominant contribution to $a_{CP}$. The constraints on
$(\ov{\rho},\ov{\eta})$ derived from $\ov{Br}$ and $a_{CP}$ are scheme
independent, of course.

\end{document}